\documentclass[twocolumn,prd,showpacs,floatfix,%
preprintnumbers,nofootinbib]{revtex4}

\usepackage{epsfig}
\usepackage{amsmath,bm}

\begin{document}

\title{Limits on a CP-violating scalar axion-nucleon interaction}

\author{Georg Raffelt}
\affiliation{Max-Planck-Institut f\"ur Physik
(Werner-Heisenberg-Institut), F\"ohringer Ring 6, 80805 M\"unchen,
Germany}

\date{9 May 2012, finalized 22 June 2012}

\begin{abstract}
Axions or similar hypothetical pseudoscalar bosons may have a small
CP-violating scalar Yukawa interaction $g_s^N$ with nucleons,
causing macroscopic monopole-dipole forces. Torsion-balance
experiments constrain $g_p^e g_s^N$, whereas $g_p^N g_s^N$ is
constrained by the depolarization rate of ultra-cold neutrons or
spin-polarized nuclei. However, the pseudoscalar couplings $g_p^{e}$
and $g_p^{N}$ are strongly constrained by stellar energy-loss
arguments and $g_s^N$ by searches for anomalous monopole-monopole
forces, together providing the most restrictive limits on $g_p^{e}
g_s^N$ and $g_p^{N} g_s^N$. The laboratory limits on $g_s^N$ are
currently the most restrictive constraints on CP-violating axion
interactions.
\end{abstract}

\preprint{MPP-2012-74}


\maketitle

%%%%%%%%%%%%%%%%%%%%%%%%%%%%%%%%%%%%%%%%%%%%%%%%%%%%%%%%%%%%%%%%%%%%%%
\section{Introduction}                               \label{sec:intro}
%%%%%%%%%%%%%%%%%%%%%%%%%%%%%%%%%%%%%%%%%%%%%%%%%%%%%%%%%%%%%%%%%%%%%%

The Peccei-Quinn mechanism for explaining the absence of
CP-violating effects in QCD leads to the prediction of axions, new
pseudoscalar bosons with a very small mass~\cite{Peccei:2006as,
Kim:2008hd}. Such particles would mediate new macroscopic forces
between spin-polarized bodies (dipole-dipole forces), which however
are hard to measure because they compete with magnetic interactions.
Monopole-dipole and monopole-monopole forces will also arise if
axions have small CP-violating scalar interactions with
nucleons~\cite{Moody:1984ba}. Axions were invented to explain the
absence of CP violation in QCD and indeed residual CP-violating
standard-model effects will be extremely small~\cite{Georgi:1986kr}.
However, new sources of CP violation may well exist and provide
neutron and nuclei electric dipole moments and CP-violating
axion-nucleon interactions with a phenomenologically interesting
magnitude~\cite{Barbieri:1996vt, Pospelov:1997uv, Pospelov:2005pr}.

A new force on macroscopic scales would be a major discovery of
fundamental importance. Precision tests of Newton's inverse square
law and of the weak equivalence principle have a long
tradition~\cite{Fischbach:1992fa, Adelberger:2009zz}. Besides
looking for new forces between bulk matter (monopole-monopole
forces), one can also look for ``unnatural parity'' monopole-dipole
forces. The hypothesis of CP violation in axion interactions
provides one motivation, but of course the measurements themselves
are agnostic of the underlying theory.

Torsion-balance experiments can look for new forces between bulk
matter and a body with polarized electrons. They are interpreted in
terms of the pseudoscalar interaction $g_p^e$ of a new boson $\phi$
(for example the axion) and the scalar interaction $g_s^N$ with
nucleons. One derives constraints on the product $g_p^e g_s^N$,
depending on the assumed range $\lambda=1/m_\phi$ of the new force.
(We always use natural units with $\hbar=c=1$.) Another class of
experiments studies the spin depolarization of nuclei or neutrons
under the influence of the surrounding bulk matter, providing limits
on the product of scalar and pseudoscalar interaction with nucleons
$g_p^N g_s^N$. Likewise, one can study the relative procession
frequencies of atoms or look for an induced magnetization in a
paramagnetic salt.

We here show that the scalar and pseudoscalar couplings are
individually constrained, leading to more restrictive limits on the
product $g_s g_p$ than provided by the current generation of
monopole-dipole force experiments. The scalar nucleon interaction
$(g_s^N)^2$ is best constrained by searches for anomalous
monopole-monopole forces. The pseudoscalar interaction $g_p^e$ is
constrained by the energy loss of white dwarfs and globular-cluster
stars, $g_p^N$ by the neutrino signal duration of SN~1987A. There
also exist direct laboratory bounds on the pseudoscalar couplings
from dipole-dipole force experiments, but the results are not yet
competitive with stellar energy-loss limits.

We juxtapose the constraints on $g_s g_p$ thus derived with those
from monopole-dipole force measurements. This comparison provides a
benchmark for the required sensitivity improvements for the direct
force experiments to enter unexplored territory in parameter space.

In Sec.~\ref{sec:astrolimits} we briefly review the astrophysical
limits on new boson interactions. In Sec.~\ref{sec:baryon} we
summarize experimental limits on the scalar nucleon interaction. In
Sec.~\ref{sec:spinbulk} we juxtapose our limits on $g_s g_p$ with
those from monopole-dipole experiments and briefly mention limits on
dipole-dipole forces in Sec.~\ref{sec:dipoledipole}. In
Sec.~\ref{sec:axions} we interpret the results for axions and
conclude in Sec.~\ref{sec:conclusions}.

%%%%%%%%%%%%%%%%%%%%%%%%%%%%%%%%%%%%%%%%%%%%%%%%%%%%%%%%%%%%%%%%%%%%%%
\section{Astrophysical limits} \label{sec:astrolimits}
%%%%%%%%%%%%%%%%%%%%%%%%%%%%%%%%%%%%%%%%%%%%%%%%%%%%%%%%%%%%%%%%%%%%%%

\subsection{Electron coupling}
\label{sec:electroncoupling}

We assume that electrons couple to a low-mass boson $\phi$ through a
derivative coupling
$(C_{e\phi}/2f_\phi)\bar\psi_e\gamma^\mu\gamma_5\psi_e\partial_\mu\phi$
where $f_\phi$ is some large energy scale, in the case of axions the
Peccei-Quinn scale $f_a$, and $C_{e\phi}$ a numerical coefficient.
This is usually equivalent to the pseudoscalar interaction $-{\rm
i}g_p^e\bar\psi_e\gamma_5\psi_e\phi$ with
$g_p^e=C_{e\phi}m_e/f_\phi$. This interaction allows for stellar
energy losses by the Compton process $\gamma+e\to e+\phi$ and
bremsstrahlung $e+Ze\to Ze+e+\phi$ \cite{Raffelt:1999tx,
Raffelt:2006cw}.

The brightness of the tip of the red-giant branch in globular
clusters constrains various cooling mechanisms of the degenerate
core before helium ignition, and in particular
reveals~\cite{Raffelt:1994ry}
\begin{equation}\label{eq:gpe-astro}
g_p^e\alt3\times10^{-13}\,.
\end{equation}
This limit pertains to particles with $m_\phi\alt 10$~keV so that
their emission is not suppressed by threshold effects.

White-dwarf cooling would be accelerated by $\phi$ emission
\cite{Raffelt:1985nj}. Isern and collaborators have found that the
white-dwarf luminosity function fits better with a small amount of
anomalous energy loss that can be interpreted in terms of $\phi$
emission with $g_p^e\sim 2\times10^{-13}$ \cite{Isern:2008nt}. The
period decrease of the pulsating white dwarf G117-B15A also favors
some amount of extra cooling~\cite{Isern:2010wz}. The interpretation
in terms of $\phi$ emission is of course speculative and we adopt
Eq.~(\ref{eq:gpe-astro}) as our nominal limit.

For completeness we mention that the scalar electron coupling can be
similarly constrained~\cite{Grifols:1986fc, Raffelt:1999tx}
\begin{equation}\label{eq:gse-astro}
g_s^e\alt1.3\times10^{-14}\,.
\end{equation}
This limit is more restrictive because the emission process does not
suffer from electron spin flip.

\subsection{Nucleon coupling}

The pseudoscalar nucleon coupling, defined analogous to the electron
coupling, allows for the bremsstrahlung process $N+N\to N+N+\phi$ in
a collapsed supernova core. However, the measured neutrino signal of
SN~1987A reveals a signal duration of some 10~s and thus excludes
excessive new energy losses~\cite{Raffelt:1987yt}. The emission rate
suffers from significant uncertainties related to dense nuclear
matter effects~\cite{Janka:1995ir} and amounts to an educated
dimensional analysis \cite{Raffelt:2006cw}. Assuming equal $\phi$
couplings to protons and neutrons one finds~\cite{Raffelt:1999tx}
\begin{equation}\label{eq:gpn-astro}
g_p^N\alt3\times10^{-10}\,.
\end{equation}
In typical axion models, the interaction with neutrons can actually
vanish.

The scalar interaction is not well constrained by this method
because nucleon velocities are relatively small. Moreover, if the
neutron and proton couplings are equal, nonrelativistic
bremsstrahlung of scalars vanishes. The most restrictive
astrophysical limit arises from the energy loss of globular-cluster
stars through the process $\gamma+{}^4{\rm He}\to {}^4{\rm
He}+\phi$~\cite{Grifols:1986fc, Raffelt:1999tx, Raffelt:1988gv}
\begin{equation}\label{eq:gsn-astro}
g_s^N\alt 0.5\times10^{-10}\,.
\end{equation}
This limit is quite restrictive because the electric charges and the
scalar nucleon couplings each add coherently.

%%%%%%%%%%%%%%%%%%%%%%%%%%%%%%%%%%%%%%%%%%%%%%%%%%%%%%%%%%%%%%%%%%%%%%
\section{Scalar Baryon Interactions}                \label{sec:baryon}
%%%%%%%%%%%%%%%%%%%%%%%%%%%%%%%%%%%%%%%%%%%%%%%%%%%%%%%%%%%%%%%%%%%%%%

We next consider a long-range Yukawa force mediated by a scalar
$\phi$ that couples with equal strength $g_s^N$ to protons and
neutrons. For small $m_\phi$, restrictive limits derive from
precision tests of Newton's inverse square law. The new Yukawa
potential is traditionally expressed as a correction to Newton's
potential in the form
\begin{equation}
V=-\frac{G_{\rm N} m_1 m_2}{r}\,\left(1+\alpha\,e^{-r/\lambda}\right)\,,
\end{equation}
where, in terms of the atomic mass unit $m_u$,
\begin{equation}
\alpha=\frac{\left(g_s^N\right)^2}{4\pi\,G_{\rm N} m_u^2}=
1.37\times10^{37}\left(g_s^N\right)^2\,.
\end{equation}
The force range is
\begin{equation}
\lambda=m_\phi^{-1}=19.73~{\rm cm}~\frac{\mu{\rm eV}}{m_\phi}\,.
\end{equation}
In the literature, one usually finds plots of the limiting $\alpha$
as a function of $\lambda$; for a recent review see
Ref.~\cite{Adelberger:2009zz}.

\begin{figure}[b]
\includegraphics[width=1.0\columnwidth]{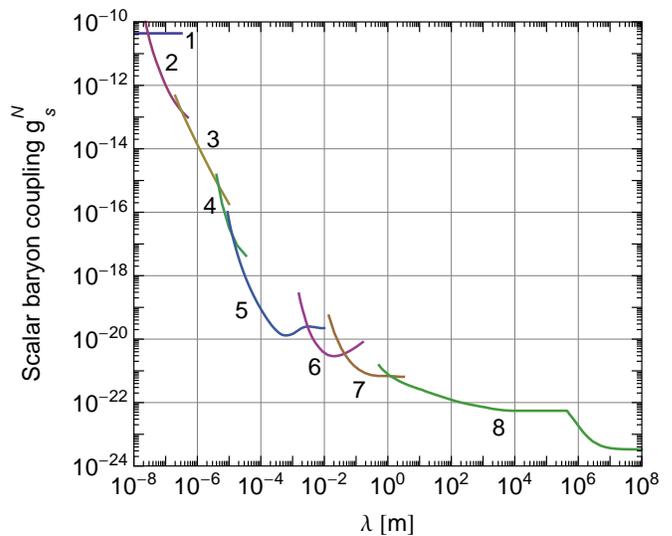}
\caption{Limits on the scalar $\phi$ coupling to
baryons. Curve~1 derives from stellar energy
loss~\cite{Grifols:1986fc, Raffelt:1999tx}. Curves~2--6 depend on
tests of Newton's inverse square law~\cite{Decca:2007jq, Sushkov:2011zz,
Geraci:2008hb, Kapner:2006si, Hoskins:1985tn}. Curves~7--8 derive from
testing the weak equivalence
principle~\cite{Smith:1999cr, Schlamminger:2007ht}.\label{fig:scalar}}
\end{figure}

New scalar interactions with nucleons can be probed in different
ways. Stellar energy-loss arguments are most effective for boson
masses so large that the interaction range is too short for
laboratory tests. Next one can search for deviations from the
inverse-square law (ISL) behavior of the overall force between
bodies. At the largest distances, tests of the weak equivalence
principle (WEP) are most effective, i.e.\ one searches for force
differences on bodies with different composition and in this way
isolates the non-gravitational part~\cite{Adelberger:2009zz}. The
results of such experiments can be interpreted in different ways,
depending on the assumed property of the new force. We only consider
scalar forces interacting with baryon number, but of course one can
go through the same arguments for other assumptions.

Following the numbers of curves in Fig.~\ref{fig:scalar}, at the
shortest distances (1)~the stellar energy-loss limit of
Eq.~(\ref{eq:gsn-astro}) beats laboratory limits. (2)~At distances
around $10^{-7}$~m, the Casimir measurements of Decca et al.\ (2007)
are most relevant~\cite{Decca:2007jq}, (3)~followed around the $\mu$m
scale by those of Sushkov et al.\ (2011) at
Yale~\cite{Sushkov:2011zz}. (4)~At the 10~$\mu$m scale, Geraci et al.
(2008) of the Stanford group have reported limits on deviations from
Newton's law using cryogenic micro-cantilevers~\cite{Geraci:2008hb}.
(5)~Torsion-balance tests of the inverse-square law conducted by the
E\"ot-Wash Collaboration (Kapner et al.\ 2007) provide the best
limits in the 10~$\mu$m--few~mm range~\cite{Kapner:2006si}. (6)~In
the cm range, the Irvine group's (Hoskins et al.\ 1985) torsion
balance inverse-square tests dominate~\cite{Hoskins:1985tn}. For
larger distances, one has to rely on tests of the equivalence
principle where we assume that $\phi$ couples only to baryon number.
(7)~In the sub-meter range, we use the E\"ot-Wash limits of Smith et
al.\ (1999) \cite{Smith:1999cr} and (8) at yet larger distances those
of Schlamminger et al.\ (2008) \cite{Schlamminger:2007ht}.

%%%%%%%%%%%%%%%%%%%%%%%%%%%%%%%%%%%%%%%%%%%%%%%%%%%%%%%%%%%%%%%%%%%%%%
\section{Monopole-Dipole forces}              \label{sec:spinbulk}
%%%%%%%%%%%%%%%%%%%%%%%%%%%%%%%%%%%%%%%%%%%%%%%%%%%%%%%%%%%%%%%%%%%%%%

\subsection{Electron-Nucleon Interaction}

\begin{figure}
\includegraphics[width=0.97\columnwidth]{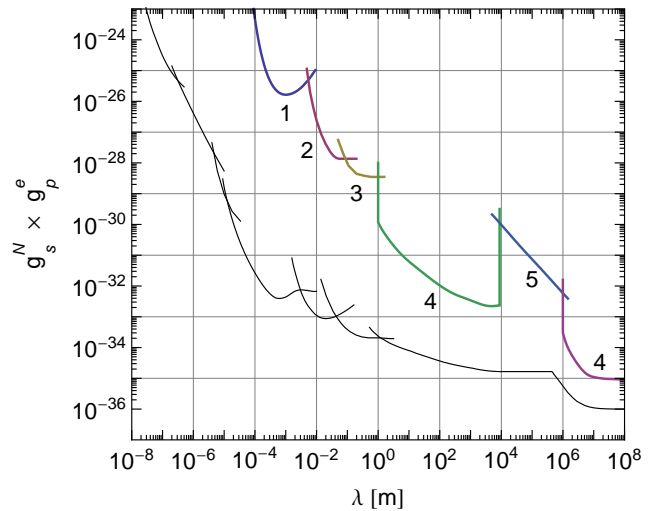}
\caption{Upper limits on $g_s^N g_p^e$. The thin black line represents the
$g_s^N$ limits of Fig.~\ref{fig:scalar} multiplied with the astrophysical
$g_p^e$ limit of Eq.~(\ref{eq:gpe-astro}). The experimental curves 1--5
constrain monopole-dipole forces~\cite{Hoedl:2011zz, Ni:1999di,
Youdin:1996dk, Heckel:2008hw, Wineland:1991zz}.\label{fig:ne}}
\end{figure}

The most restrictive limit on $g_s^N g_p^e$ arises from the
long-range force limits on $g_s^N$ shown in Fig.~\ref{fig:scalar} and
the astrophysical limit on $g_p^e$ limit of Eq.~(\ref{eq:gpe-astro}).
We show the product as the lower thin black line in
Fig.~\ref{fig:ne}. We recall that for deriving the limits on $g_s^N$
it was assumed that the scalar coupling applies only to baryon
number, whereas the pseudoscalar coupling applies to electrons.

Constraints from searches for monopole-dipole forces with torsion
pendulums using polarized electrons are shown in Fig.~\ref{fig:ne}.
(1)~The most recent constraints in the mm range were derived by
Hoedl et al.\ (2011) with a dedicated apparatus~\cite{Hoedl:2011zz}.
(2)~In the cm range, the best constraints are from the older
measurements of the Tsing Hua University group (Ni et al.\ 1999)
using a paramagnetic salt in a rotating copper
mass~\cite{Ni:1999di}. (3)~At 10~cm we show constraints derived by
Youdin et al.\ (1996) by comparing the relative precession
frequencies of Hg and Cs magnetometers as a function of the position
of two 475 kg lead masses with respect to an applied magnetic
field~\cite{Youdin:1996dk}. (4)~In the meter-range and above, the
torsion pendulum measurements of the E\"ot-Wash Collaboration
(Heckel et al.\ 2008) provide the most restrictive
limits~\cite{Heckel:2008hw}, except in a gap at 10--1000~km.
(5)~Here we fall back on stored-ion spectroscopy (Wineland et al.\
1991) \cite{Wineland:1991zz}.

\subsection{Nucleon-Nucleon Interaction}

\begin{figure}
\includegraphics[width=0.99\columnwidth]{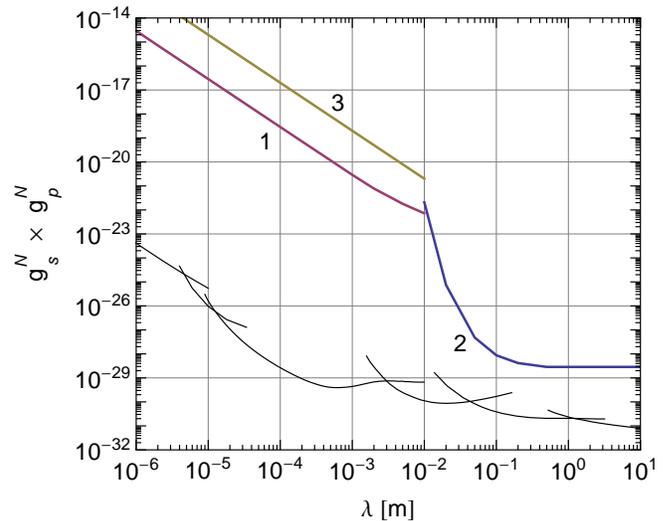}
\caption{Upper limits on $g_s^N g_p^N$. The thin black line represents the
$g_s^N$ limits of Fig.~\ref{fig:scalar} multiplied with the SN~1987A limit on
$g_p^N$ of Eq.~(\ref{eq:gpn-astro}). Curve~1 is the experimental limit
from $^3$He depolarization~\cite{Petukhov:2010dn}, curve~2
from mercury precession~\cite{Youdin:1996dk}, and curve~3 from ultra-cold
neutrons~\cite{Serebrov:2009pa}.\label{fig:nn}}
\end{figure}

The most restrictive limit on $g_s^N g_p^N$ also arises from the
long-range force limits of Fig.~\ref{fig:scalar} together with the
SN~1987A limit on the pseudoscalar coupling of
Eq.~(\ref{eq:gpn-astro}). We show the product as a thin black line
in Fig.~\ref{fig:nn}.

The most restrictive direct experimental limit at short distances
arises from measurements of the depolarization of the $^3$He
nucleus. We show the limits of Petukhov et al. (2010)
\cite{Petukhov:2010dn} as curve 1 in Fig.~\ref{fig:nn}. (2)~In the
cm range and above, the precession of Hg and Cs (Youdin et al.\
1996) provide the best limits~\cite{Youdin:1996dk}. (3)~We also show
constraints from the precession and depolarization of ultra-cold
neutrons (Serebrov et al. 2010) \cite{Serebrov:2009pa}.

Constraints from gravitational bound states of ultra-cold neutrons
\cite{Baessler:2006vm} are at the moment not competitive, but may
hold significant promise for the future \cite{Abele:2011}.

%%%%%%%%%%%%%%%%%%%%%%%%%%%%%%%%%%%%%%%%%%%%%%%%%%%%%%%%%%%%%%%%%%%%%%
\section{Dipole-Dipole Forces} \label{sec:dipoledipole}
%%%%%%%%%%%%%%%%%%%%%%%%%%%%%%%%%%%%%%%%%%%%%%%%%%%%%%%%%%%%%%%%%%%%%%

Dipole-dipole forces have been constrained by laboratory
experiments, although the results are less restrictive than the
corresponding astrophysical limits. For the pseudoscalar neutron
coupling one finds $g_p^n<0.85\times10^{-4}$ for $m\alt 10^{-7}$~eV
based on a K--$^3$He comagnetometer~\cite{Vasilakis:2008yn}. For the
pseudoscalar electron coupling, the most recent E\"ot-Wash torsion
balance spin-spin experiment yields $g_p^e< 3\times10^{-8}$ for
$m\alt 10^{-6}$~eV \cite{Adelberger:2012}.

%%%%%%%%%%%%%%%%%%%%%%%%%%%%%%%%%%%%%%%%%%%%%%%%%%%%%%%%%%%%%%%%%%%%%%
\section{Axion interpretation}                      \label{sec:axions}
%%%%%%%%%%%%%%%%%%%%%%%%%%%%%%%%%%%%%%%%%%%%%%%%%%%%%%%%%%%%%%%%%%%%%%

These limits on the various scalar and pseudoscalar couplings of a
hypothetical low-mass boson can be interpreted specifically in terms
of QCD axions where the interaction strengths and mass are closely
correlated apart from model-dependent numerical factors.

\begin{figure}
\includegraphics[width=0.84\columnwidth]{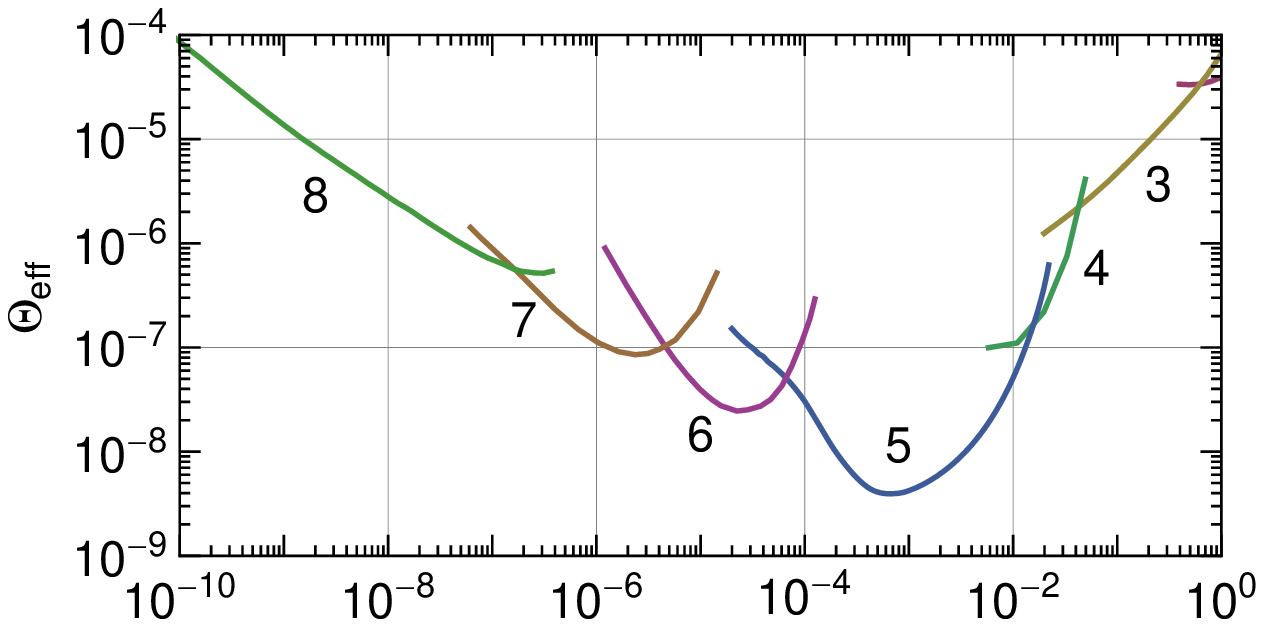}
\includegraphics[width=0.84\columnwidth]{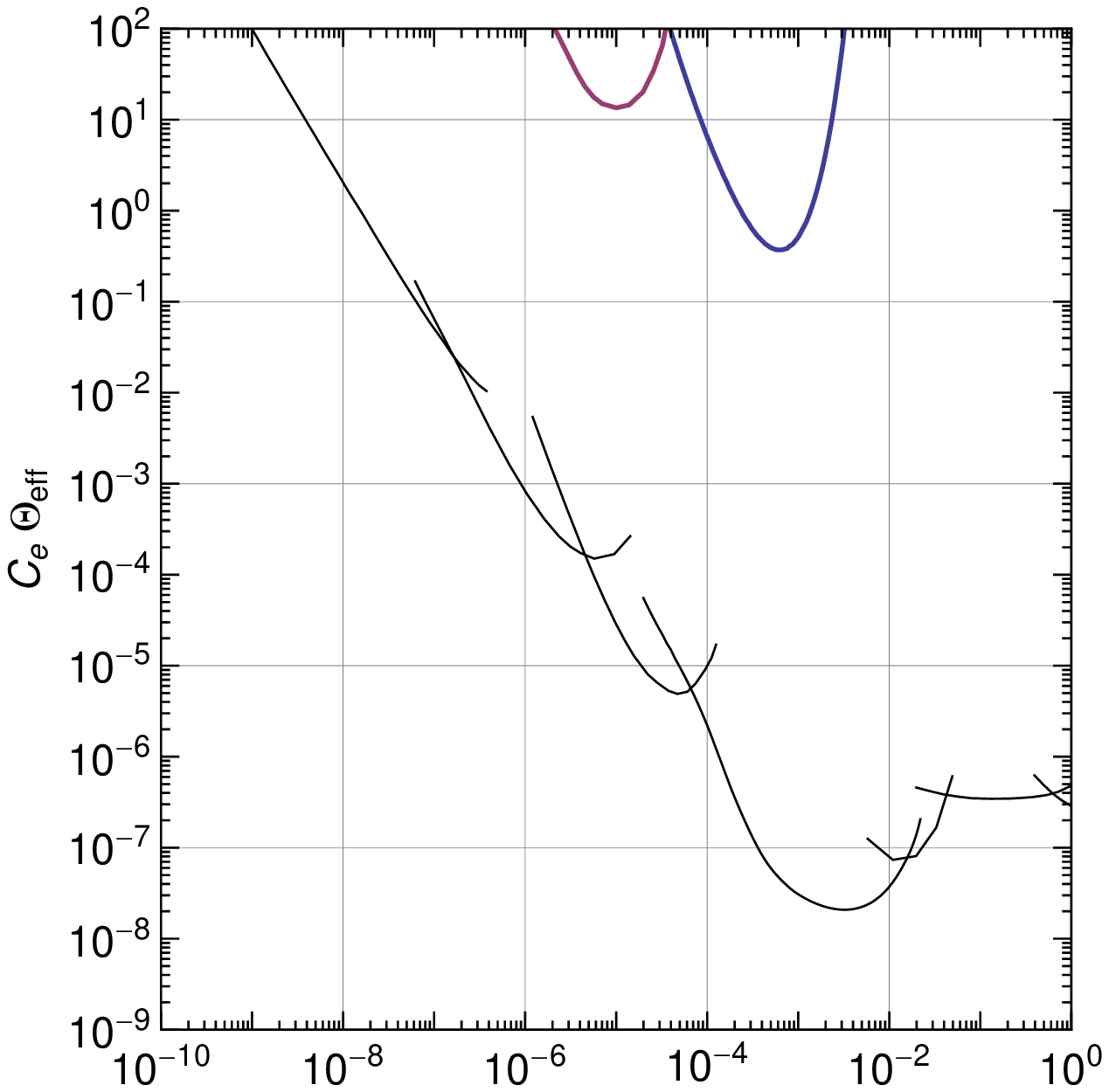}
\includegraphics[width=0.84\columnwidth]{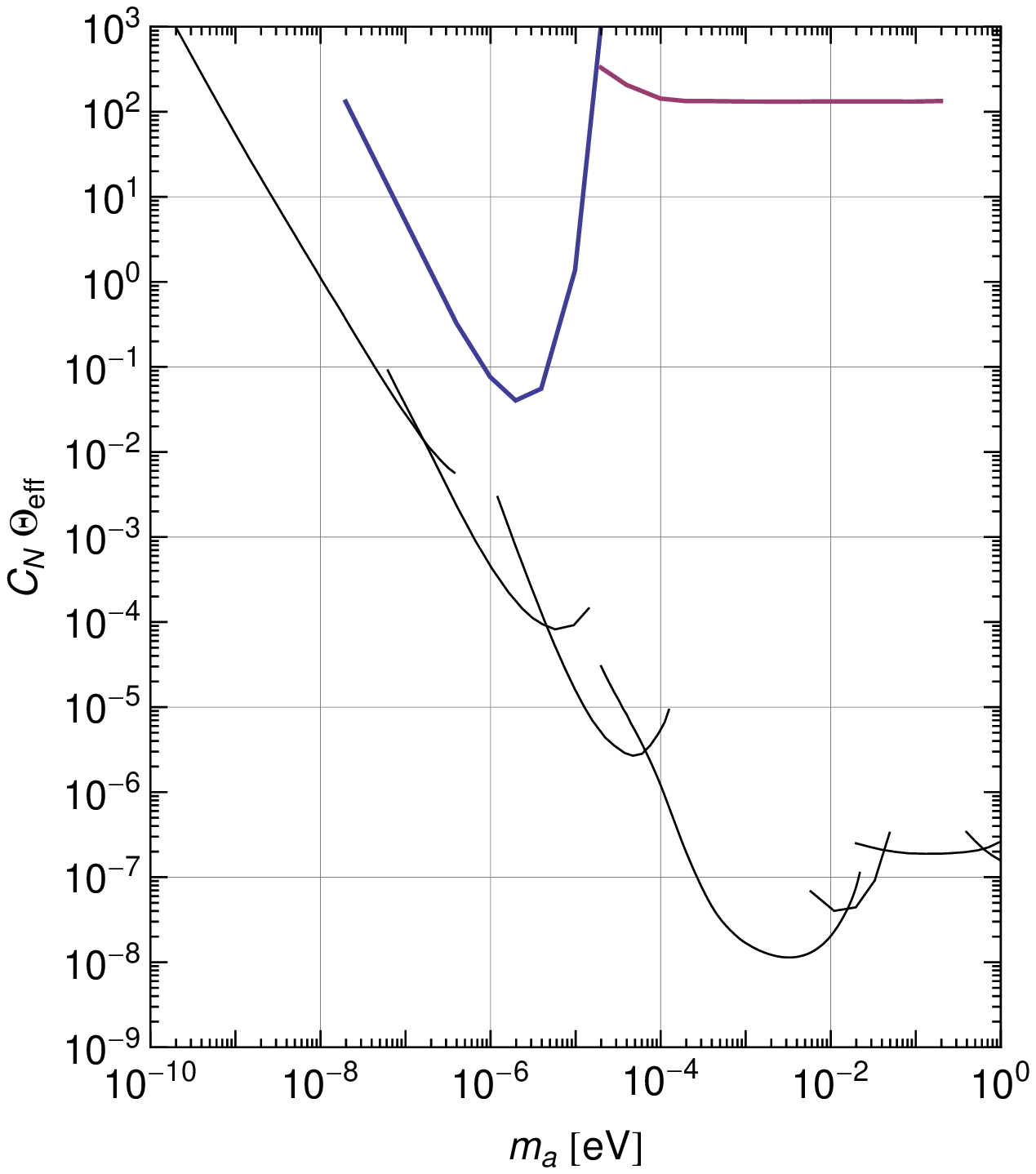}
\caption{Long-range force limits translated to the effective
CP-violating axion parameter $\Theta_{\rm eff}$ using:
{\it Top}: $g_s^N$ of Fig.~\ref{fig:scalar} and Eq.~(\ref{eq:theta}).
{\it Middle}: $g_s^N g_p^e$ of Fig.~\ref{fig:ne} and Eq.~(\ref{eq:ne}).
{\it Bottom}: $g_s^N g_p^N$ of Fig.~\ref{fig:nn} and Eq.~(\ref{eq:nn}).
\label{fig:theta}}
\end{figure}

One characteristic of axions is the relation $m_a f_a\sim m_\pi
f_\pi$ between their mass $m_a$, decay constant $f_a$, pion mass
$m_\pi=135$~MeV and pion decay constant $f_\pi=92$~MeV. A
CP-violating scalar interaction can be expressed
as~\cite{Moody:1984ba, Pospelov:1997uv}
\begin{equation}\label{eq:theta}
g_s^N\sim\Theta_{\rm eff}\,\frac{f_\pi}{f_a}\sim
\Theta_{\rm eff}\,\frac{m_a}{m_\pi}\,,
\end{equation}
where $\Theta_{\rm eff}$ measures CP-violating effects. Taking this
relation as defining $\Theta_{\rm eff}$ we show in
Fig.~\ref{fig:theta} (top) the $g_s^N$ limits translated into limits
on $\Theta_{\rm eff}$ as function of $m_a$.

Axions with $m_a$ exceeding about 1~eV are excluded by cosmological
hot dark matter bounds~\cite{Hannestad:2010yi} and $m_a$ exceeding
about 10~meV by the energy loss of SN~1987A. The meV range would be
favored by anomalous white-dwarf cooling
(Sec.~\ref{sec:electroncoupling}). It is interesting that
Fig.~\ref{fig:theta} (top) shows greatest sensitivity at this
``axion meV frontier'' \cite{Raffelt:2011ft}. However, even in this
range the $\Theta_{\rm eff}$ sensitivity is far from realistic
values because limits on neutron and nuclear electric dipole moments
imply $\Theta_{\rm eff}\alt10^{-11}$ \cite{Pospelov:1997uv,
Pospelov:2005pr}.

The pseudoscalar axion-electron interaction is $g_p^e=C_e
m_e/f_a\sim C_e (m_e/f_\pi)(m_a/m_\pi)$, where $C_e$ is a
model-dependent coefficient. Overall we therefore have
\begin{equation}\label{eq:ne}
g_s^N g_p^e\sim
\Theta_{\rm eff}\,C_e\,\frac{m_e}{f_\pi}\left(\frac{m_a}{m_\pi}\right)^2\,.
\end{equation}
Using this relation we translate the $g_s^N g_p^e$ limits of
Fig.~\ref{fig:ne} into $C_e\Theta_{\rm eff}$ and show the result in
Fig.~\ref{fig:theta} (middle).

Likewise, the pseudoscalar axion-nucleon interaction is $g_p^N=C_N
m_N/f_a\sim C_N (m_N/f_\pi)(m_a/m_\pi)$ so that
\begin{equation}\label{eq:nn}
g_s^N g_p^N\sim
\Theta_{\rm eff}\,C_N\,\frac{m_N}{f_\pi}\left(\frac{m_a}{m_\pi}\right)^2\,.
\end{equation}
Translating the $g_s^N g_p^N$ limits of Fig.~\ref{fig:nn} into
limits on $C_N\Theta_{\rm eff}$ leads to Fig.~\ref{fig:theta}
(bottom).

For the moment any of these limits are far from the
phenomenologically interesting range. In a more detailed analysis,
one should include differences of the axion coupling to protons and
neutrons.

%%%%%%%%%%%%%%%%%%%%%%%%%%%%%%%%%%%%%%%%%%%%%%%%%%%%%%%%%%%%%%%%%%%%%%
\section{Conclusions}                          \label{sec:conclusions}
%%%%%%%%%%%%%%%%%%%%%%%%%%%%%%%%%%%%%%%%%%%%%%%%%%%%%%%%%%%%%%%%%%%%%%

We have interpreted existing laboratory limits on anomalous
monopole-monopole forces into limits on the scalar interaction
$g_s^N$ of a new low-mass boson $\phi$ with baryons. We have
combined them with stellar energy-loss limits on the pseudoscalar
$\phi$ coupling with electrons $g_p^e$ and nucleons $g_p^N$ and have
derived the most restrictive limits yet on the products $g_s^Ng_p^e$
and $g_s^Ng_p^N$. These constraints are more restrictive than
laboratory searches for anomalous monopole-dipole forces. Of course,
pure laboratory searches remain of utmost importance, especially if
they can eventually overtake the astrophysical results.

%%%%%%%%%%%%%%%%%%%%%%%%%%%%%%%%%%%%%%%%%%%%%%%%%%%%%%%%%%%%%%%%%%%%%%
\section*{Acknowledgements} %%%%%%%%%%%%%%%%%%%%%%%%%%%%%%%%%%%%%%%%%%
%%%%%%%%%%%%%%%%%%%%%%%%%%%%%%%%%%%%%%%%%%%%%%%%%%%%%%%%%%%%%%%%%%%%%%

I thank Hartmut Abele, Peter Fierlinger and John Ellis for
discussions at the Symposium ``Symmetries and Phases of the
Universe'' (February 2012) that motivated this work, Eric Adelberger
for discussions at the workshop ``Vistas in Axion Physics'' (April
2012), and Maxim Pospelov and Seth Hoedl for thoughtful comments on
the manuscript. Partial support from the Deutsche
Forschungsgemeinschaft grant EXC-153 and from the European Union FP7
ITN INVISIBLES (Marie Curie Actions, PITN-GA-2011-289442) is
acknowledged.

%%%%%%%%%%%%%%%%%%%%%%%%%%%%%%%%%%%%%%%%%%%%%%%%%%%%%%%%%%%%%%%%%%%%%%
%%%  Bibliography  %%%%%%%%%%%%%%%%%%%%%%%%%%%%%%%%%%%%%%%%%%%%%%%%%%%
%%%%%%%%%%%%%%%%%%%%%%%%%%%%%%%%%%%%%%%%%%%%%%%%%%%%%%%%%%%%%%%%%%%%%%

%%%%%%%%%%%%%%%%%%%%%%%%%%%%%%%%%%%%%%%%%%%%%%%%%%%%%%%%%%%%%%%%%%%%%%
\end{document}